# Phase Imaging with Computational Specificity (PICS) for measuring dry mass changes in sub-cellular compartments

**Author List:** Mikhail E. Kandel[1,2,†], Yuchen R. He[1,2,†], Young Jae Lee[1,3], Taylor Hsuan-Yu Chen[1,4], Kathryn Michele Sullivan[4], Onur Aydin[5], M Taher A. Saif[4,5], Hyunjoon Kong[1,4,6,7], Nahil Sobh[1*], Gabriel Popescu[1,2,4*]

1. Beckman Institute, University of Illinois at Urbana-Champaign, Urbana, IL, USA.
2. Department of Electrical and Computer Engineering, University of Illinois at Urbana-Champaign, Urbana, IL, USA.
3. Neuroscience Program, University of Illinois at Urbana-Champaign, Urbana, IL, USA.
4. Department of Bioengineering, University of Illinois at Urbana-Champaign, Urbana, IL, USA.
5. Department of Mechanical Science and Engineering, University of Illinois at Urbana-Champaign, Urbana, IL, USA.
6. Chemical and Biomolecular Engineering, University of Illinois at Urbana-Champaign, Urbana, IL, USA.
7. Carl Woese Institute for Genomic Biology, University of Illinois at Urbana-Champaign, Urbana, IL, USA.

[†] Equal contributions

*Correspondence to:

Nahil Sobh, 4039 Beckman Institute, 405 North Mathews Ave, Urbana, Illinois 61801, (217) 244-1176, sobh@illinois.edu

Gabriel Popescu, 4055 Beckman Institute, 405 North Mathews Ave, Urbana, Illinois 61801, (217) 333-4840, gpopescu@illinois.edu

**Abstract:**

Due to its specificity, fluorescence microscopy (FM) has become a quintessential imaging tool in cell biology. However, photobleaching, phototoxicity, and related artifacts continue to limit FM's utility. Recently, it has been shown that artificial intelligence (AI) can transform one form of contrast into another. We present PICS, a combination of quantitative phase imaging and AI, which provides information about unlabeled live cells with high specificity. Our imaging system allows for automatic training, while inference is built into the acquisition software and runs in real-time. Applying the computed fluorescence maps back to the QPI data, we measured the growth of both nuclei and cytoplasm independently, over many days, without loss of viability. Using a QPI method that suppresses multiple scattering, we measured the dry mass content of individual cell nuclei within spheroids. In its current implementation, PICS offers a versatile quantitative technique for continuous simultaneous monitoring of individual cellular components in biological applications where long-term label-free imaging is desirable.

Fluorescence microscopy has been the most common imaging tool for studying cellular biology[1]. Fluorescence signals, whether intrinsic or extrinsic, allow the investigator to study particular structures in the biospecimen with high specificity[2]. However, this important quality comes at an expensive price: chemical toxicity and phototoxicity disturb and may kill a live cell[3,4], while *photobleaching* limits the extent of the investigation window[5]. Breakthroughs in genetic engineering led to the family of green fluorescent proteins, which today are broadly used in live cells with reduced toxicity[6]. In addition, current research efforts are dedicated to reducing photobleaching by various methods, including oxygen scavenging[7] and replacing traditional fluorophores with quantum dots[8].

Microscopy with intrinsic contrast preceded fluorescence labeling by more than two centuries[9]. Advanced forms of label-free imaging, such as phase-contrast microscopy, developed in the 1930's[10], and differential interference contrast, in the 1950's[11], extended the capability of imaging transparent specimens, including live cells. However, the lack of chemical specificity and inability to inform on underlying mechanisms has relegated these modalities to routine tasks, such as visual inspection of tissue cultures. Thus, fluorescence microscopy has remained a necessity for in-depth biology.

Recently, quantitative phase imaging (QPI) has advanced label-free microscopy with its ability to extract quantitative parameters (cell dry mass, cell mass transport, cell tomography, nanoscale morphology, topography, pathology markers, etc.) from unlabeled cells and tissues[12]. As a result, QPI can extract structure and dynamics information from live cells without photodamage or photobleaching[13-19]. However, in the absence of labels, QPI cannot easily identify particular structures in the cell as the label-free image lacks specificity.

In a parallel development, within the past few years, in part due to the continuous decline of computing power cost, development of frameworks for dataflow representation as well as a steep increase in data generation, deep learning techniques have been translating from consumer to scientific applications[20-24]. For example, it has been shown that AI can map one form of contrast into another, a concept coined as *image-to-image translation*[25-28]. Significantly, it has been demonstrated that a neural network can predict the mapping of a stain or fluorescence marker from label-free images as input[29,30].

Inspired by this prior work, we present a new microscopy concept, referred to as phase imaging with computational specificity (PICS), in which the process of learning is automatic and retrieving computational specificity is part of the acquisition software, performed in real-time (Fig. 1). We applied deep learning to QPI data, generated by SLIM (spatial light interference microscopy)[31-34] and GLIM (gradient light interference microscopy)[35,36]. These methods are white-light and common-path and, thus, provide high spatial and temporal sensitivity[37-44]. Because they are add-ons to existing microscopes and compatible with the fluorescence channels, these methods provide simultaneous phase and fluorescence images from the same field of view. As a result, the training data necessary for deep learning is generated automatically, without the need for manual annotation. This new type of microscopy can potentially replace some commonly used tags and stains and eliminate the inconveniences associated with chemical tagging. We demonstrate this idea with various fluorescence tags and diverse cell types, at different magnifications, on different QPI systems. We show that combining QPI and computational specificity allows us to quantify the growth of subcellular components (*e.g.* nucleus vs cytoplasm) over many cell cycles, nondestructively. Finally, using

GLIM, we imaged spheroids and demonstrated that PICS can perform single-cell nucleus identification even in such turbid structures.

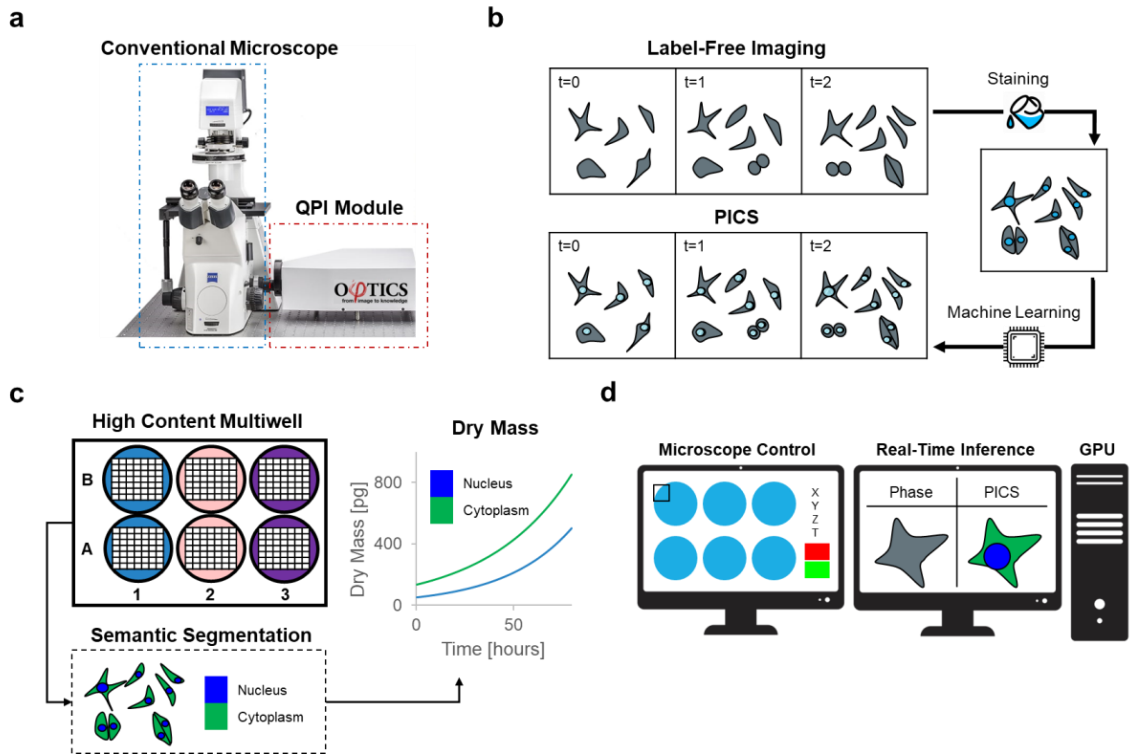

**Figure 1. PICS method for label-free measurements of compartment-specific cellular dry mass. a,** We upgrade a conventional transmitted light microscope with a quantitative phase imaging add-on module. **b,** To avoid the intrinsic toxicity of fluorescent stains, we develop a two-step protocol imaging protocol where label-free images are recorded followed by fixation and staining. From the toxic stain recorded at the end of the experiment, we train a neural network capable of digitally staining the time-lapse sequence, thus enabling time-lapse imaging of otherwise toxic stains. **c,** The digital stain is used to introduce specificity to label-free imaging by providing a semantic segmentation map labeling the components of the cell. From the time-lapse sequence, we calculate organelle-specific dry mass doubling times, in this case, the rates of growth for the nucleus and cytoplasm. **d,** The PICS method is integrated into a fully automated plate reading instrument capable of displaying the machine learning results in real-time.

PICS advances the field of AI-enhanced imaging in several ways. First, PICS performs *automatic registration* by recording both QPI and fluorescence microscopy of the same field of view, on the same camera. The two imaging channels are integrated seamlessly by our software that controls both the QPI modules, fluorescence light path, and scanning stage. The PICS instrument can scan a large field of view, e.g., entire microscope slides, or multi-well plates, as needed. Second, PICS can achieve fluorescence channel multiplexing by automatically training on multiple fluorophores but performing inference on a single phase image. Because PICS uses intrinsic contrast images as input, which benefit from strong signals, it provides an order of magnitude improvement in acquisition rate compared to traditional fluorescence microscopy. Third, PICS performs *real-time inference,* because we incorporated the AI code into the live acquisition software. The computational inference is faster than the image acquisition rate in SLIM and GLIM, which is up to 15 frames per second, thus, we add specificity without noticeable delay. To the microscope user, it would be difficult to state whether the live image originates in a fluorophore or the computer GPU. Fourth, using the specificity maps obtained by computation, we exploit the QPI channel to compute the dry mass density image associated with the particular subcellular structures. For example, using this procedure, we demonstrated a previously unachievable task: the measurement of growth curves of cell nuclei vs. cytoplasm over several days, nondestructively. Fifth, using a QPI method dedicated to imaging 3D cellular systems (GLIM), we add subcellular specificity to turbid structures such as spheroids.

**Results**

### *PICS Method*

The PICS methodology is outlined in Fig. 1. We use an inverted microscope (Axio Observer Z1, Zeiss) equipped with a QPI module (CellVista SLIM Pro and CellVista GLIM Pro,

Phi Optics, Inc.). The microscope is programmed to acquire both QPI and fluorescence images of fixed, tagged cells (Fig. 2b). Once the microscope "learned" the new fluorophore, PICS can perform inference on the live, never labeled cells. Due to the absence of chemical toxicity and photobleaching, as well as the low power of the white light illumination, PICS can perform dynamic imaging over arbitrary time scales, from milliseconds to weeks, without cell viability concerns. Simultaneous experiments involving multi-well plates can be performed to assay the growth and proliferation of cells of specific cellular compartments (Fig. 1c). Finally, the inference is implemented within the QPI acquisition time, such that PICS performs in real-time (Fig. 1d).

PICS combines quantitative measurements of the object's scattering potential with fluorescence microscopy. The essentials of the QPI optics and computation are shown in Fig 2. Fig. 2a illustrates the optical path of the GLIM system used for most of the QPI results in this work (for completeness, SLIM is described in Supplementary Fig. 1). The GLIM module controls the phase between the two interfering fields outputted by a DIC microscope, as described in Supplementary Note 1[35]. We acquired four intensity images corresponding to phase shifts incremented in steps of $\pi/2$ and combined these to obtain a quantitative phase gradient map (Fig. 2b). This gradient is integrated using a Hilbert transform method, as described in Supplementary Fig. 2 and Supplementary Note 2. The same camera records fluorescence images via epi-illumination providing a straightforward way to combine the fluorescence and phase images. Fig. 2b illustrates the acquired images consisting of two fluorescence channels (cell nuclei and membrane in this case) and GLIM.

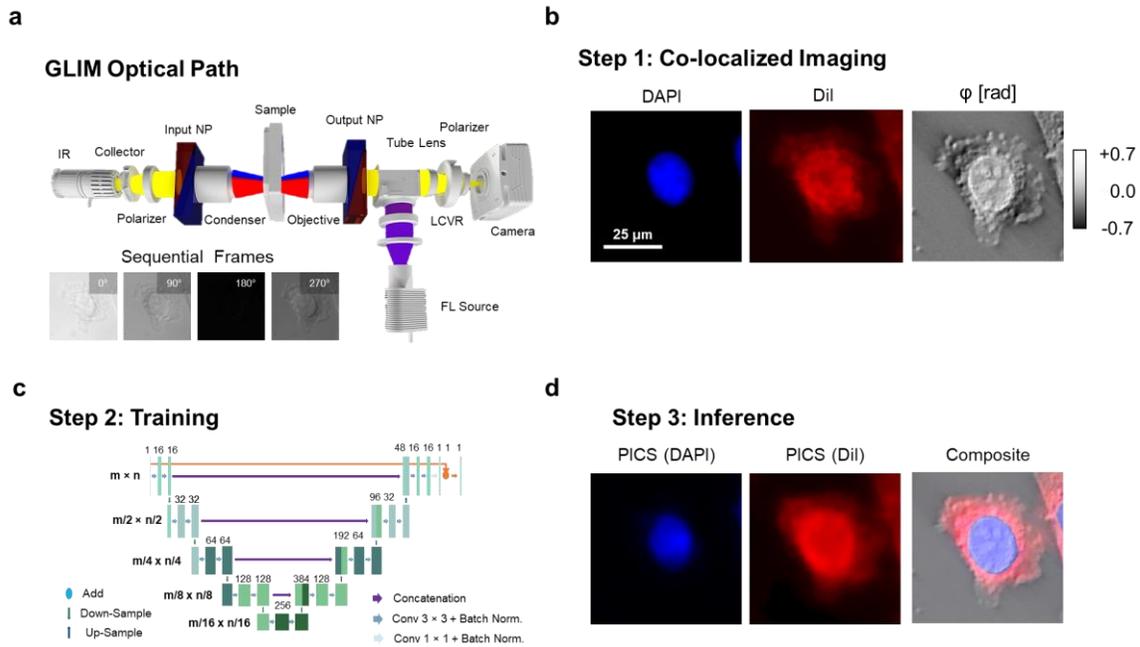

**Figure 2. The stain location is "learned" from co-localized phase and fluorescent images.**
**a,** Quantitative phase images are acquired with a compact Gradient Light Interference Microscopy (GLIM) module that attaches to the output port of a differential interference contrast microscope (DIC/Nomarksi). By using a liquid crystal variable retarder (LCVR) the instrument introduces controlled phase shifts between the orthogonal polarizations in DIC. GLIM images are the result of a four-frame reconstruction process to retrieve the phase associated with a differential interference contrast microscope. Insets show a zoomed portion of the field of view at 0°, 90°, 180°, 270° phase shifts. **b,** The same light path is used for reflected light fluorescence imaging, providing a straightforward way to co-localize fluorescence and phase images. In this work, we focused on two popular stains used to assay the nucleus and cell body (DAPI and DiI). To recover the phase-shift associated with the object's scattering potential, we remove the shear artifact associated with the DIC field by performing integration using a Hilbert transform to obtain φ, the phase shift measured along the DIC shear direction. Zoomed portion of a field of view showing a typical SW620 cell (20x/0.8). **c,** Next, to learn the mapping between the label-free and stained image, we train a U-Net style deep convolutional neural network. **d,** Once this model is trained we can perform real-time interference and rendering to obtain the equivalent fluorescence signal (PICS) directly from the label-free image.

We use these co-localized image pairs to train a deep convolutional neural network to map the label-free phase images to the fluorescence data. For deep learning, we used a variant of U-Net by introducing three modifications. First, following the work by Google[45], we added batch normalization layers before all the activation layers, which helped accelerate the training. Second, we greatly reduced the number of parameters in our network by changing the number of feature maps in each layer of the network to a quarter of what was proposed in the original paper. This change greatly reduced GPU memory usage and improved inference time, without loss of performance. Our modified U-Net model used approximately 1.9 million parameters, while the original architecture had over 30 million parameters. Based on the training results, we believe that 1.9 million parameters are sufficient to approximate the mapping from phase images to fluorescence images. Third, we utilized the advantage of residual learning[46] with the hypothesis that it is easier for the models to approximate the mapping from phase images to the difference between phase images and fluorescence images. Thus, we implemented an add operation between the input and the output of the last convolutional block to generate the final prediction. We noticed that this change enabled us to have much better performance under the same training conditions. The modified network architecture is shown in Fig. 2c (orange connection) and described in more detail in Supplementary Fig. 3. Fig. 2d shows the result of the inference. To measure the performance of PICS under various conditions, we applied this procedure across different image resolutions, fluorophores, and cell lines, using both SLIM and GLIM (see Supplementary Fig. 4). Our training dataset consisted of three most in-focus images of each unique field of view, spaced 2-3 depths of field apart. This approach serves as a natural form of data augmentation. We also realized that the amount of data needed for satisfying results differ from task to task (see Supplementary Table 1).

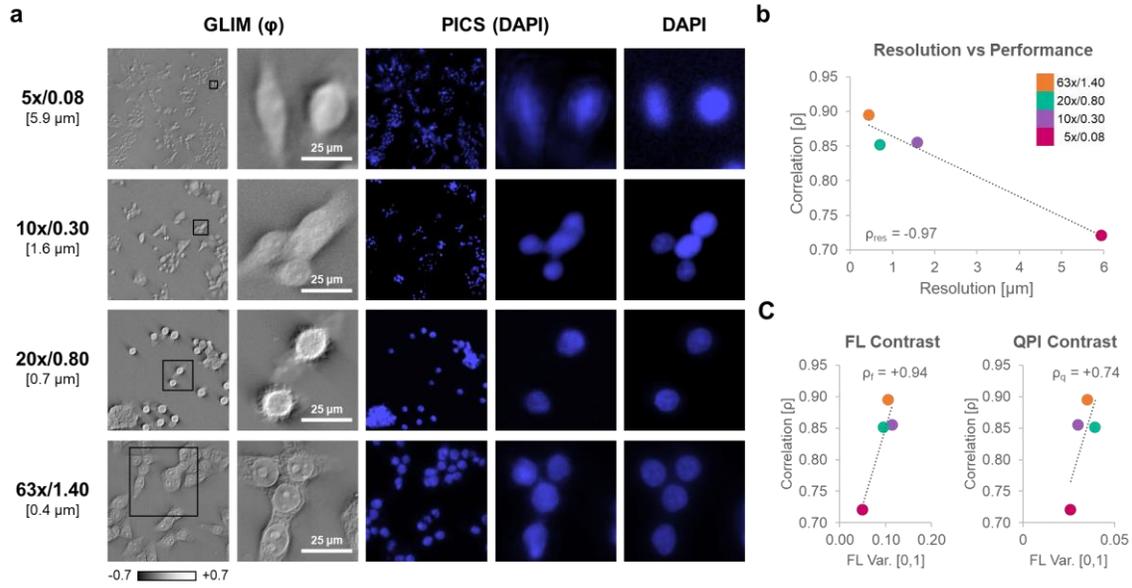

**Figure 3: The PICS method is applicable across microscope objectives and resolutions. a,** To investigate the effect of resolution on performance we run a computational experiment where we train our network on SW cells acquired at different resolutions. To control for training sample size, we selected the most images shared between all sets. We note that the performance of the network improves with more data or training epochs. Even at low resolutions (10x, 1.6 μm resolution), we achieved adequate performance. **b,** We use ρ, the correlation between the actual and the digital fluorescent signals over the entire set, as a quality metric. There appears to be a clear relationship (Pearson correlation, $\rho_{res}$ = -0.97) between resolution (better objectives) and performance. **c,** The origin of the relationship between resolution and performance is attributable to differences in contrast recorded by these objectives. We find that the $\rho_f$ coefficient between the variance of training scaled fluorescence images and quality metric ρ is statistically significant while the correlation between the variance of the training scaled phase images $\rho_q$ is weaker. Overall, these results suggest that the relationship between performance associated with resolution can be largely attributed to better overall contrast, especially for the fluorescent signal, rather than directly due to resolving capabilities.

### *Effects of resolution on PICS performance*

To understand the performance of our approach we conducted a series of computational experiments where we held the training time and quantities of training pairs constant and vary the objectives used for imaging. Figure 3 shows in detail how the resolution of QPI impacts the values of the Pearson correlation coefficient that quantifies the match between the computationally predicted and actual fluorescence images. Remarkably, even for a 5x/0.08NA objective, the Pearson correlation is above 72%. Figure 3c presents the effect of the image contrast upon the performance. As expected, higher contrast, *i.e.,* spatial variance, of both the fluorescence and QPI yields better performance. Note that, for a fair comparison, we kept the number of epochs constant across all resolutions and contrast, which somewhat limited the network performance (see Supplemental Table 1). A comparison between quantitative phase imaging, standard contrast enhancement techniques (DIC), and bright field microscopy is presented in (Supplementary Fig. 5). The data indicate that the addition of the interferometric hardware to decouple phase and amplitude information improves the performance of the AI algorithm. Furthermore, PICS provides a uniform and consistent stain. Supplementary Fig. 5 highlights a staining defect that PICS was able to correct.

### *Training data set considerations for PICS*

To study the relationship between the number of training pairs vs. prediction accuracy, we conducted a second series of computational experiments where we varied the size of the data set while keeping other training parameters constant. As shown in Supplementary Fig. 6, we found that high fidelity digital stains can be generated from as few as 20 image pairs (roughly 500 SW cells) corresponding to five minutes of training time. When the performance of our procedure is cross-validated by training on a subset of the data (Supplementary Fig. 6a), we found that certain images were dominant for training (Supplementary Fig. 7). In other words, we found that certain folds converge faster than others. Importantly, neural networks that performed well during training on small data sets (Supplementary Fig. 6a), also performed well when being validated on larger, unseen data sets (Supplementary Fig. 6b). For example, the five minutes used to train a neural network from 20 pairs is well below the time typically needed to stain the cells (Supplementary Fig. 6c).

Supplementary Fig. 7 contains a summary of the 57 networks trained for this work.

## *Time-lapse PICS of adherent cells*

To illustrate the value of specificity multiplexing, i.e., inferring multiple stains on the same cell, we acquired simultaneous PICS images of both the cell nucleus and membrane (Figs. 4 & 5). Supplementary Video 1 shows the data acquisition procedure. After training, the inference model was integrated into the acquisition software for real-time operation with both SLIM and GLIM (Supplementary Video 2-3). Supplementary Figs. 8 & 9 describe the typical acquisition sequence and operation of the instrument. We note, that in general, fluorescence tags required an order of magnitude more exposure time than the QPI frames, implying that our plate reader achieves higher throughput while maintaining specificity. This effect is amplified when separate exposures are used for individual fluorophores.

Because of the nondestructive nature of PICS, we can apply it to monitor cells over extended periods, of many days, without a noticeable loss in cell viability. This important aspect is emphasized in Fig. 4 and Supplementary Video 4. To perform a high content cell growth screening assay, *unlabeled* SW480 and SW620 cells were imaged over seven days and PICS predicted both DAPI (nucleus) and DiI (cell membrane) fluorophores. The density of the cell culture increased significantly over the seven days, a sign that cells continued their multiplication throughout the imaging. Note that, in principle, PICS can multiplex numerous stain predictions simultaneously, as training can be performed on an arbitrary number of fluorophores for the same cell type. The only price paid is computational time, as each inference channel adds, ~65 ms to the real-time inference. The computation time for one stain is completely masked by the acquisition process and multiple networks can be evaluated in parallel on separate GPUs.

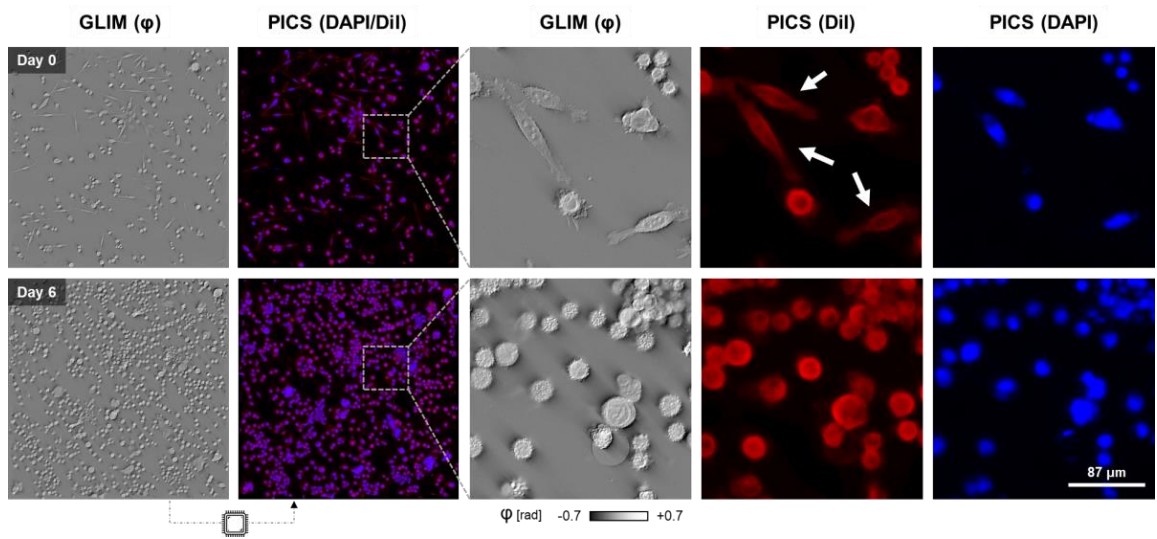

**Figure 4: Time-lapse PICS of unstained cells**. To demonstrate time-lapse imaging and high content screening capabilities, we seeded a multiwell with three distinct concentrations of SW cells (20x/0.8). These conditions were imaged over the course of a week by acquiring mosaic tiles consisting of a 2.5 mm$^2$ square area in each well using a 20x/0.8 objective. The machine learning classifier, trained at the final time point after paraformaldehyde fixation, is applied to the previously unseen sequence to yield a DiI and DAPI equivalent image. Interestingly, the neural network was able to correctly reproduce the DiI stain on more elongated fibroblast-like cells, even though few such cells are present when the training data was acquired (white arrows).

## *Cell growth measurements of sub-cellular compartments*

We used PICS-DiI to generate a binary mask (Fig. 5, Supplementary Fig. 10), which, when applied to the QPI images, yields the dry mass of the entire cell. Similarly, PICS-DAPI allows us to obtain the nuclear dry mass. Thus, we can independently and dynamically monitor the dry mass content of the cytoplasm and nucleus. This capability is illustrated in Figs. 5B and C, where an individual cell is followed through mitosis. It is known that the nuclear-cytoplasmic ratio (NCR) is a controlling factor in embryogenesis[47] and a prognosis marker in various types of cancer[48,49]. Figures 5d-f show the specific growth curves for a large cell field of view, consisting of a mosaic of covering a 2.5 mm$^2$ portion of a multiwell. Figure 5g illustrates the behavior of the confluence factor (defined as a fraction of the total area occupied by the cells) in time. Not surprisingly, as the confluence increases, the growth saturates due to contact inhibition[50]. In Supplementary Fig. 11, we repeat this imaging protocol, demonstrating that the median dry mass of the nuclei remains stable over time, while the area distinguishes between different cell lines (SW480 vs SW620). Interestingly, in this cell co-culture, while the metastatic cells (SW620) have smaller nuclei, the total dry mass is similar to that of SW480. Note that we used the same neural network for both cell lines.

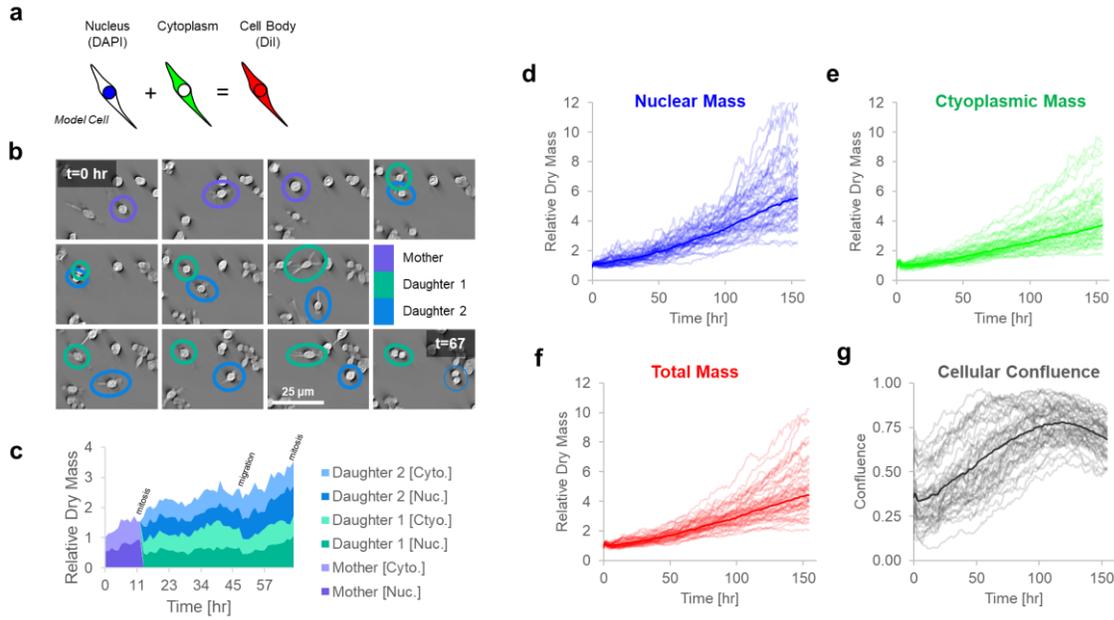

**Figure 5: Tracking dry mass changes in cellular compartments using PICS. a,** The DiI and DAPI stains are specific to the cell body and nucleus, respectively. The difference between the two areas produces a semantic segmentation map that distinguishes between the nuclear and non-nuclear content of the cell (cytoplasm). **b,** Throughout the experiment, we observe cellular growth and proliferation with cells often traveling a substantial distance between division events. **c,** Using the semantic segmentation map we can track the nuclear and cytoplasmic dry mass. We find that nuclear and cytoplasmic dry mass steadily increase until mitosis, with some loss of dry mass due to cellular migration. **d-g,** Semantic segmentation maps enable us to track the nuclear and cytoplasmic dry mass and area over 155 hours. The dark curve represents the median of the growth rate across forty-nine fields of view (lighter curves). The dry mass and area are normalized by the average measured value from the first six hours. In this experiment we observe that total nuclear dry mass grows faster than total cytoplasmic mass, providing further evidence that cells can divide without growing. As the cells reach optimal confluence (t≈114 hours), we observe a decrease in the growth rate of nuclear mass, although less difference in cytoplasmic dry mass growth.

## *PICS of Spheroids*

GLIM has been developed recently in our laboratory to extend QPI applications to thicker, strongly scattering structures, such as embryos[35], spheroids, and acute brain slices[36]. GLIM improves image quality by suppressing artifacts due to multiple scattering and provides a quantitative method to assay cellular dry mass. To showcase this capability, we imaged 20 spheroids using GLIM equipped with a 63x/1.4NA objective. Each spheroid was imaged in depth over an 85 μm range, sampled in steps of 80 nm, with each field of view measuring 170x170 μm$^2$. At each z-position, epi-fluorescence imaging was also performed to reveal the DAPI-stained nuclei (Fig. 6a). Following the same training procedure as before, we found that PICS can infer the nuclear map with high accuracy. Specifically, we constructed a binary mask using PICS and DAPI images and compared the fraction of mass found inside the two masks. Thus, Fig. 6b shows that the average error between inferring nuclear dry mass based on the DAPI vs. PICS mask is 4%.

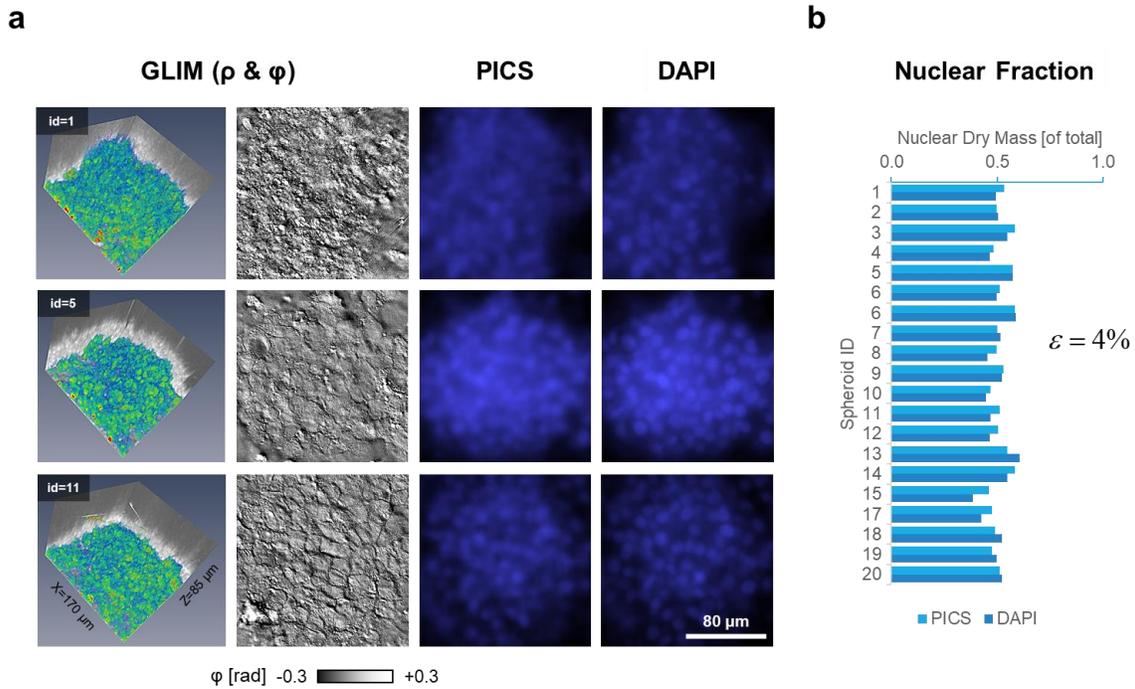

**Figure 6: PICS for digitally staining 3D cellular systems. a,** Representative images of PICS applied to HepG2 spheroids (63x/1.4, 170 μm x 170 μm x ~85 μm). The scattering potential, GLIM (ρ) was recovered, as in the original GLIM paper, by nonlinear filtering where the absolute value of the phase map is displayed on a log scale. PICS and DAPI insets were bilaterally filtered to improve contrast. **b,** To compare the performance of PICS to conventional DAPI staining we constructed a semantic segmentation map by thresholding the DAPI and PICS image and calculating the dry mass within this map. When comparing total dry mass across twenty samples, we find the average percentage change between the PICS and DAPI images to be 4%.

**Discussion and Outlook**

PICS uses artificial intelligence to boost the capability of QPI. PICS exploits the unique capabilities of SLIM and GLIM, whereby the QPI and fluorescence images can be obtained by the same camera, without the need for complex image registration. As a result, annotation, which normally represents a bottleneck for AI, is performed automatically, with no manual intervention. In principle, the number of fluorescent channels that PICS can predict is virtually unlimited. Our approach is to collect training data at the end of the experiment on fixed cells, effectively training for each cell type and magnification. This training is performed only once and ensures that performance is optimal. Once the training dataset is stored on the computer, the microscope user benefits from a virtual stain that can be used indefinitely. The network inference requires a mere 65 ms per frame, which is faster than the image acquisition for both SLIM and GLIM. This inference time is also approximately one order of magnitude shorter than the typical exposure in our fluorescence imaging. As a result, the specificity map is displayed in real-time and overlaid with the QPI phase map.

The main benefit of PICS over regular fluorescence is the fact that computation is, of course, nondestructive, while at the same time, QPI yields quantitative information. Furthermore, the QPI data used as input is obtained using low levels of light, which has low phototoxicity. Thus, we demonstrated time-lapse imaging of live cells over a week while maintaining cell viability intact and a high level of specificity for cellular compartments. This capability is particularly valuable when studying cell growth, which remains an insufficiently understood phenomenon[51]. In particular, we showed that by multiplexing specificity for cell nuclei and lipid bilayers, PICS can simultaneously assay nuclear, cytoplasm, and total cell growth over many cell cycles. In this way, by training on fixed cells at the end of the experiment, PICS mimics fluorescence stains (such as DAPI) that are otherwise incompatible with live-cell imaging. The approach of learning stains from fixed cells for live-cell imaging presents many opportunities. For example, there is a pressing need for developing live-cell imaging techniques capable of reproducing stains that are associated with protein expression (antibody) or membrane permeability (cell viability[52]) as these stains require fixation.

Interestingly, we found that by decoupling the amplitude and phase information, QPI images outperform their underlying modalities (phase contrast, DIC) in AI tasks (Supplementary Fig. 5). This capability is showcased in GLIM, which provides high-contrast imaging of thick tissues by suppressing multiple scattering, enabling us to achieve subcellular specificity in optically turbid spheroids. We foresee a range of applications in this area, including viability assays in spheroids subjected to various treatments[53].

Finally, because PICS can be implemented as a hardware add-on module to an existing microscope, the threshold for adaption in the field is low. The automatic training procedure allows the user to easily replace the chemical makers in their studies. The real-time inference gives instantaneous feedback about the sample, which keeps the user experience virtually unchanged, while operating at an improved throughput and reduced toxicity.
As shown in Supplementary Fig. 12, PICS reveals that the quantitative phase image contains vastly more information than the fluorescence counterpart. In a broader context, PICS illustrates a paradigm shift in microscopy, where the resurgence of intrinsic contrast imaging is empowered by recent advances in deep-learning methods to gain specificty.

## Methods

### *Acquisition procedure*

With respect to Ref.[54], our current software was designed as a "frontend" with acquisition dialogs to generate lists of events that are then processed by a "backend". The principal changes to the backend involved instrumenting TensorRT (NVIDIA) for real-time inference, while the frontend changes involved developing a graphic interface to facilitate plate-reader style imaging. PICS images are processed following the scheme shown in Supplementary Fig. 8. Each PICS image is the result of an acquisition sequence that collates four label-free intensity images into a phase map. The sequence begins by introducing a phase shift on the modulator ("Modulation") followed by camera exposure and readout. In GLIM, the phase shift is introduced by a liquid crystal variable retarder (Thorlabs), which takes approximately 70 ms to fully stabilize. In SLIM a ring pattern is written on the modulator and 20 ms is allowed for the crystal to stabilize (Meadowlark, XY Series). Next, four such intensity images are collated to reconstruct the phase map ("Phase Retrieval"). In GLIM, the image is integrated (6 ms) and in SLIM we remove the phase-contrast halo artifact (25 ms). The phase map is then passed into a deep convolution neural network based on the U-Net architecture to produce a synthetic stain (65 ms). Finally, the two images are rendered as an overlay with the digital stain superimposed on the phase image (5 ms). In the "live" operating mode used for finding the sample and testing the network performance, a PICS image is produced for every intensity frame. Under typical operation, the rate-limiting factor is the speed of image acquisition rather than computation time. As a point of comparison, the two-channel fluorescence images used to train PICS required approximately 1000 ms of integration time making PICS approximately 15 times faster.

*Real-time PICS*

The PICS system uses an optimized version of the U-Net deep convolutional neural architecture to translate the quantitative phase map into a fluorescence one. To achieve real-time inference, we use TensorRT (NVIDIA) which automatically tunes the network for the specific network and GPU pairings[55]. In the process of this work, we found that the TensorRT was unable to parse standard machine learning interchange formats such as ONNX and instead developed a script to convert the model from TensorFlow (Google) to the optimized TensorRT inference engine (NVIDIA). In short, this script converts the weights learned by TensorFlow to match the format supported by TensorRT. The network was instrumented layer-by-layer using the TensorRT's C++ API. In addition to performance gains, TensorRT can operate directly on GPU memory, avoiding redundant data copies.

The PICS inference framework is designed to account for differences between magnification and camera frame size. Differences in magnification are accounted for by scaling the input image to the networks' required pixel size using NVIDIA's Performance Primitives library. While TensorRT is fast, the network-tuning is performed online and can take a non-negligible time to initialize (30 seconds). To avoid tuning the network for each camera sensor size, we construct an optimized network for the largest image size and extend smaller images by mirror padding. Further, to avoid the edge artifacts typical of deep convolutional neural networks, a 32-pixel mirror pad is performed for all inferences.

*Multi-well plate reader operation*

Large samples, such as the multiwell plates (12.7 x 85 cm), used in this work are difficult to image due to a small but significant tilt introducing during sample placement. In this work, we compensate for sample tilt by developing a graphic user interface for plate reader applications that present each well as a 3D tomogram (Fig. S9). Tilts are controlled by adding focus points which are used to construct a Delaunay triangulation to interpolate the plane of best focus across the mosaic tiles[54]. As the glass bottom of a multiwell is flat, we found that most tilts are linear, and good results can be achieved by focusing on the four points at the corners of the well. In addition to specifying focus points, and controlling the dimensions of the acquisition, the interface configures the microscope for multichannel acquisition with fluorescence microscopy presented alongside phase imaging. The interface presents phase imaging specific features such as modulator stabilization time and variable exposure for intensity frames.

## *Training the neural networks*

We chose to use the U-Net architecture[56], which effectively captures the broad features typical of quantitative phase images. Networks were built using TensorFlow and Keras, with training performed on a variety of computers including workstations (NVIDIA GTX 1080 & GTX 2080) as well as isolated compute nodes (HAL, NCSA, 4x NVIDIA V100). Supplementary Table 1 contains a summary of the 57 networks trained in this study. No transfer learning was performed in this work. All networks were trained with the adaptive moment estimator (ADAM) against a mean squared error optimization criterion. Phase and fluorescence microscopy images, $I(x, y)$, were normalized for machine learning as

$$I_{\text{ml input}}(x, y) = \text{med}\left(0, \frac{I(x, y) - \rho_{\min}}{\rho_{\max} - \rho_{\min}}, 1\right) \quad [1]$$

where $\rho_{\min}$ and $\rho_{\max}$ are the minimum, and maximum pixel values across the entire training set, and med is a pixel-wise median filter designed to bring the values within the range [0,1]. Spatio-temporal broadband quantitative phase images exhibit strong sectioning and defocus effects. To address focus related issues, images were acquired as a tomographic stack. The Haar wavelet criterion from[54] was used to select the three most in-focus images for each mosaic tile.

### *Cell culture*

The SW480 and SW620 pairing is a popular model for cancer progression as the cells were harvested from the tumor of the same patient before and after a metastasis event[57]. Cells obtained from ATCC were grown in Leibovitz's L-15 media with 10% FBS and 1% pen-strep at atmospheric $CO_2$. Mixed SW cells were plated at a 1:1 ratio at approximately 30% confluence. The fluorescent lipophilic dye, DiI is used to stain the cell membrane. The application of the dye was adapted and modified from established protocol from the Thermofisher website. After the passage, mixed SW cells were allowed for two days to attach and grow in the well plate. When the cells reach the desired confluence, we prepared the staining medium by mixing 5uL of DiI labeling solution into the 1 mL of normal growth medium. We aspirated off all the previous medium on the well plate and pipetted staining medium to cover all the surface of the well plate for 20 minutes at 37°C. After the incubation, we drained off the staining medium and washed the cells with warmed regular growth medium three times every 10 minutes. Cells were then fixed with freshly prepared 4% paraformaldehyde (PFA) for 15 minutes and washed with PBS two times before DAPI staining. To visualize the nucleus, DAPI was used for the experiment and DAPI solution prepared with 10uL DAPI in 10mL PBS. The cells were incubated in DAPI solution for 10 minutes and washed three times with PBS before imaging.

Time-lapse microscopy was performed eight hours after platting and the slower growth rate at the start of the experiment can be attributed to the cells being in the "lag-phase" of the cell cycle[58]. The growth characteristics are consistent between experiments, suggesting that they are a constant behavior of our particular subclone.

CHO cells are commonly used for mass production of mammalian proteins[59]. CHO cells (ATCC) were cultured in Ham's F-12 with 10% FBS and 1% pen-strep under 5% $CO_2$.

HepG2 spheroids represent a kind of liver cancer that is popular for high-throughput toxicity assays. Spheroids were cultured on a glass-bottom dish as indicated in[60], which formed spheroids at sufficiently high density. To perform an experiment typical for high content screening, we plated cells on a poly-D-lysine coated multiwell.

As plastic affects the differential interference contrast, all cells imaged in this work were cultured on glass-bottom dishes covered with a DIC specific glass lid (P06-20-1.5-N, L001, Cellvis). While the glass lid can be avoided in SLIM imaging, using a plastic lid with GLIM will result in a total loss of interferometric contrast. All cells except the spheroids were grown on poly-D-lysine treated glass.

*Time-lapse microscopy*

To illustrate the nondestructive specificity associated with PICS, we performed automated time-lapse microscopy for a week. This procedure was repeated twice (Fig. 4 & 5 and again in Supplementary Fig. 11). For Figs 4-5, three conditions of cancer cells were plated in a 2 x 3 multiwell at 5 different depths. For each well, we acquire a 7 by 7 mosaic grid. This procedure is repeated for every well, with a sample taken every sixty-eight minutes. As the sample was imaged in a temperature-controlled incubator, we did not observe appreciable focus drift during the week. The resulting sequence consisted of 202,860 GLIM images, which were assembled into a mosaic by software developed in house[54]. After the experiment completed, the cells were fixed, stained with DiI and DAPI and imaged to produce a training corpus for AI. To illustrate the value of the dry mass and area, a similar procedure was applied for Supplementary Fig. 11 except SW480 and SW620 were not mixed.

**Data availability statement:**

The data that support the findings of this study are available upon reasonable request.

**Availability of computer code and algorithms:**

The code and computer algorithms that support the findings of this study are available from the corresponding author upon reasonable request.

**Acknowledgments:**

We thank Dr. Catalin Chiritescu & Taha Anwar at Phi Optics for ongoing maintenance and software development of the Cell Vista microscopes used in this work. We also thank Volodymyr Kindratenko and Dawei Mu for providing access to the supercomputer used to train the neural networks (HAL Cluster, NCSA). Frozen vials of CHO cells were provided by Huimin Zhao's group at UIUC.

This work is supported by NSF 0939511 (T.S., J.K., G.P.), R01GM129709 (G.P.), R01 CA238191 (G.P.), R43GM133280-01 (G.P.). M.E.K. and M.K.S are supported by a fellowship from MBM (NSF, NRT-UtB, 1735252). This work utilizes resources supported by the National Science Foundation's Major Research Instrumentation program, grant #1725729, as well as the University of Illinois at Urbana-Champaign.

**Contributions:**

M.E.K. designed and performed imaging experiments. Y.R.H. and N.S. developed the AI model and trained the neural networks. M.E.K. & Y.R.H. instrumented the real-time inference. Y.J.L. & T.H.C. cultured and stained the cells. K.M.S. & H.K. provided spheroids. O.A. and T.A.S. provided SW cells. M.E.K. analyzed the data. G.P., M.E.K., Y.R.H. wrote the manuscript. N.S. supervised the AI work. G.P. supervised the project.

**Conflict of Interest:**

G.P. has a financial interest in Phi Optics, Inc., a company developing quantitative phase imaging technology for materials and life science applications. The remaining authors declare no competing interests.

Supplementary information for

**Phase Imaging with Computational Specificity (PICS)**

**for measuring dry mass changes in sub-cellular compartments**

**Supplementary Note 1: Gradient Light Interference Microscopy**

To show that PICS is not dependent on a particular QPI method, we used both SLIM and GLIM. GLIM is implemented as an upgrade to a conventional DIC microscope[1,2] (Fig. 1a and Fig. 2a). To reduce photodamage and multiple scattering, we used a broadband infrared source (780 nm). The sample is illuminated by two slightly shifted fields originating from a Nomarski prism. The sample is imaged by an objective with an integrated Nomarski prism, which recombines these fields and undoes the effect of the input prism. To measure the difference in phase between the two fields, we modify the optical path by introducing a liquid crystal variable retarder (LCVR, Thorlabs) between the camera and output polarizer. The LCVR enables us to control the phase shift between the two polarizations outputted by the DIC microscope. In our instruments, we record four images corresponding to $\pi/2$ phase shifts between the two beams (Fig. 2a), which lets us recover, uniquely, the phase shift associated with the DIC microscope[1]. The resulting image resembles a derivative of the phase map associated with the object as it is based on differences in phase between neighboring points. This image is then integrated using a 1D Hilbert transform as noted in the next section. While this approach can be extended to multiple shear directions (as in [3]), here we used only 1D integration as the phase-shifting components can be located completely outside our microscope, at the expense of certain streak artifacts in the integrated image. For GLIM, we used a 20x/0.8 NA objective giving us a sampling of roughly 0.3 microns per pixel, compared to the diffraction spot of 0.7 microns. To obtain an optimal resolution, all GLIM images were acquired with a fully open condenser ($NA_c$=0.55).

**Supplementary Note 2: Phase integration using the Hilbert transform**

The phase image in GLIM is the result of interfering two laterally offset or "sheared" beams. The intensity measured at the detector resembles[2],

$$I_n(\mathbf{r}) = I(\mathbf{r}) + I(\mathbf{r}+\delta_x) + 2\sqrt{I(\mathbf{r})I(\mathbf{r}+\delta_x)}\cos\left[\phi(\mathbf{r}+\delta_x) - \phi(\mathbf{r}) + \varepsilon_n\right] \quad [1]$$

where $\Delta\phi_x = \phi(\mathbf{r}+\delta_x) - \phi(\mathbf{r}) \approx \Delta\phi_x \delta_x$ is the gradient of the phase map. When the modulator is cycled using the liquid crystal variable retarder, $\varepsilon_n \simeq \frac{\pi}{2}n$ we obtain the phase shift $\Delta\phi_x$ which is the derivative of the phase along the contrast direction, $x$, scaled by the shear $\delta_x$.

To obtain the true phase map and remove the shading effect, we perform a Hilbert transform [4] along the contrast direction, which performs the following Fourier filter (Supplementary Fig. 2a),

$$\phi_x(\mathbf{k}) = \left(-i\,\text{sgn}(k_x)/\delta_x\right)\Delta\phi_x(\mathbf{k}) \quad [2]$$

where $\mathbf{k}$ is the wavevector, and $\phi_x(\mathbf{k})$ is the integrated image along the contrast direction and sgn is the signum function. As shown in [5], this operation approximates an integral, which can be implemented as a Wiener filter [6],

$$\phi_x(\mathbf{k}) = \frac{-i}{k_x + l_{\text{reg}}}\Delta\phi_x(\mathbf{k}) \quad [3]$$

where $l_{\text{reg}}$ is the regularization constant. We note that our approach is a regularized version of the Wiener filtering method when $k_x$ is large, which is also the frequency range at which the system operates with partially coherent illumination[7].

To demonstrate the ability of the Hilbert transform to recover topographic information we imaged a 3 μm polystyrene bead embedded in immersion oil. We found that the results are in good agreement with the expected phase shift (Supplementary Fig. 2b)

$$\begin{aligned}\phi_{peak} &= \tfrac{2\pi d}{\lambda}\left(n_{object} - n_{media}\right) \\ &\approx \tfrac{2\pi(3.000)}{0.780}(1.579 - 1.518) \\ &\approx 1.48 \text{ RAD}\end{aligned} \quad [4]$$

The distortion orthogonal to the contrast-bearing axis (null space of the system transfer operator) bears little significance for live cell measurements, as cell shape and growth does not have a preferential direction.

**Supplementary Note 3: Spatial Light Interference Microscopy**

SLIM upgrades a phase-contrast microscope[8] in a similar way to how GLIM upgrades DIC. In short, SLIM uses a spatial light modulator matched to the back focal plane of the objective to control the phase shift between the incident and scattered components of the optical field. Four such phase-contrast like frames are recorded to recover the phase between the two fields (Supplementary Fig. 1). Next, the total phase is obtained by estimating the phase shift of the transmitted component and compensating for the objective attenuation[9]. Finally, the "halo" associated with phase-contrast imaging is corrected by a non-linear Hilbert transform-based approach[5].

While SLIM has higher sensitivity[10], the GLIM illumination path performs better in strongly scattering samples and dense well plates. In strongly scattering samples, the incident light, which acts as the reference field in SLIM, vanishes exponentially[1]. In dense microplates, the transmitted light path is distorted by the meniscus or blocked by high walls[2]. GLIM and SLIM images were acquired with three different Axio Observer Z1 microscopes.

While in this work we focus on our own SLIM and GLIM methods, we expect PICS to be applicable to other modalities, especially where fluorescence can be easily overlaid with quantitative phase images[11-15].

**Supplementary Figure 1**

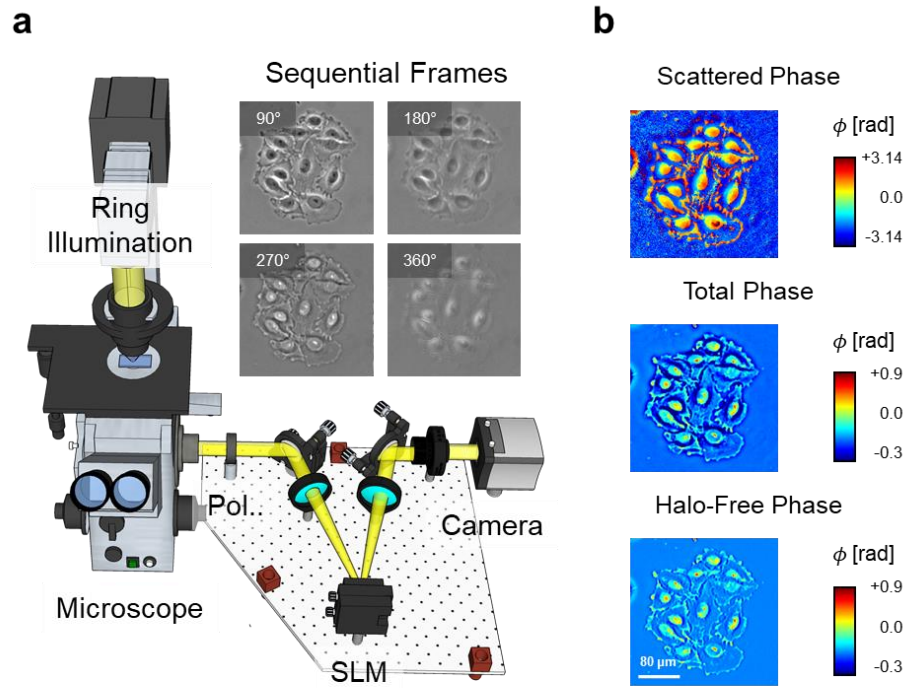

**Supplementary Figure 1: Spatial Light Interference Microscopy. a,** The ring illumination is matched to the objective's back focal plane and the mask on the spatial light modulator (SLM), effectively resulting in a phase-contrast microscope with a variable retardance ring. Four frames are recorded, corresponding to increments of 90 degrees introduced by the SLM. **b,** SLIM image reconstruction and the halo-removed SLIM image.

**Supplementary Figure 2**

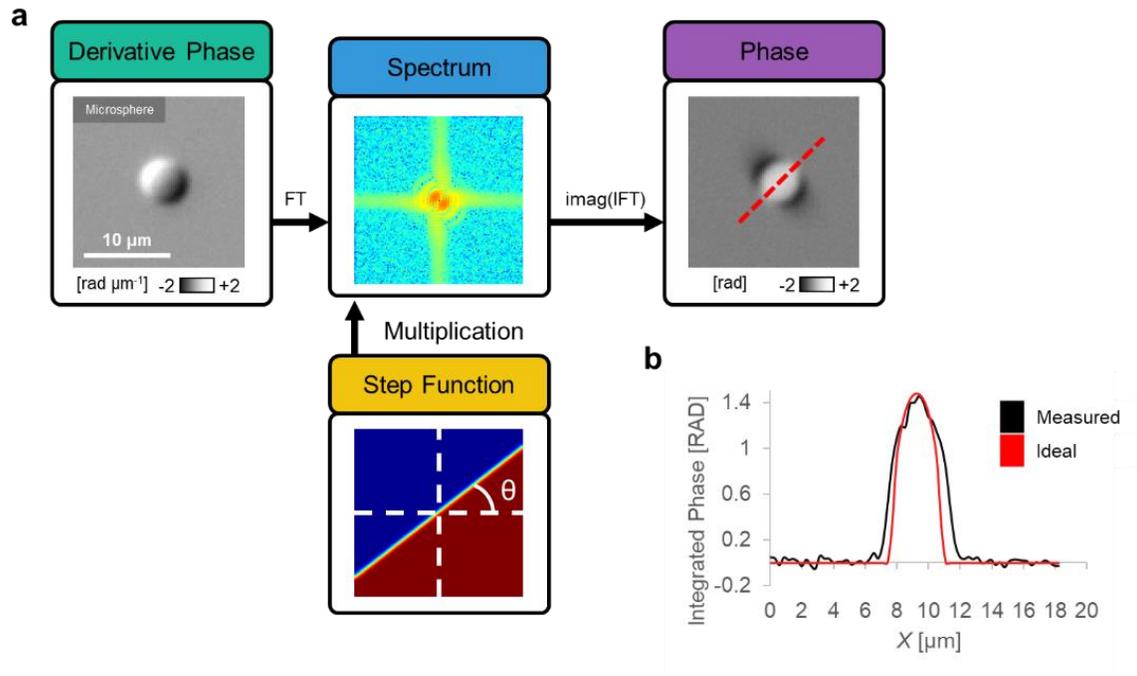

**Supplementary Figure 2: GLIM images are integrated with the Hilbert transform. a,** The Hilbert transform along the direction of the shear (θ) is performed by multiplying by a step function in the frequency domain. The imaginary portion of the inverse transform yields the integrated image. **b,** The integrated phase is in good agreement with the theoretical profile of the bead (shown on the red dashed line).

**Supplementary Figure 3**

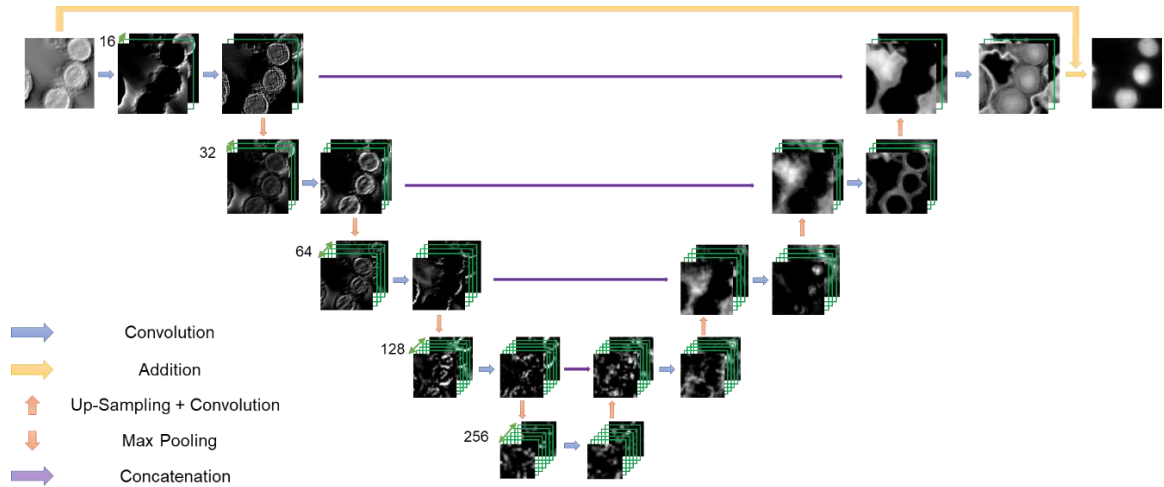

**Supplementary Figure 3: Neural Network for Phase to Fluorescence Mapping.** We modified the U-Net architecture, with batch normalization before all the activation layers and reduced the number of filters compared to the original implementation. To illustrate the evaluation of the PICS-DAPI neural network for a typical cell, we show the flow of data after applying the operations in each layer. Of particular note is the ability of the U-Net architecture to make use of both textures inside the cell (leftmost, first layers) and spatial information such as the edges around the cellular nucleolus (bottom layers).

**Supplementary Figure 4**

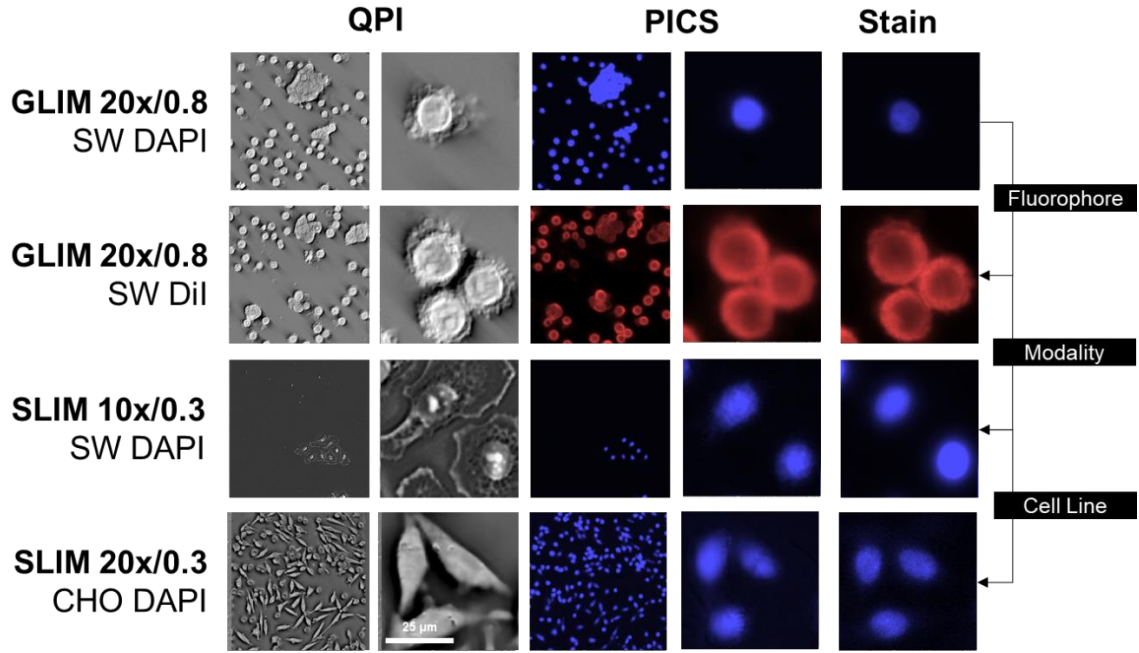

**Supplementary Figure 4: PICS method is applicable across stains, modalities, and cell-lines.** To investigate the performance of our method in various conditions, we trained separate deep convolutional neural networks on several samples and quantified their performance. As a performance metric, we compare Pearson's correlation ($\rho$) between the actual fluorescence image ("Stain") and the computationally inferred image ("PICS"). The technique is equally applicable to other QPI modalities, such as SLIM and other cell types such as CHO or a mixed culture of SW480 and SW620.

**Supplementary Figure 5**

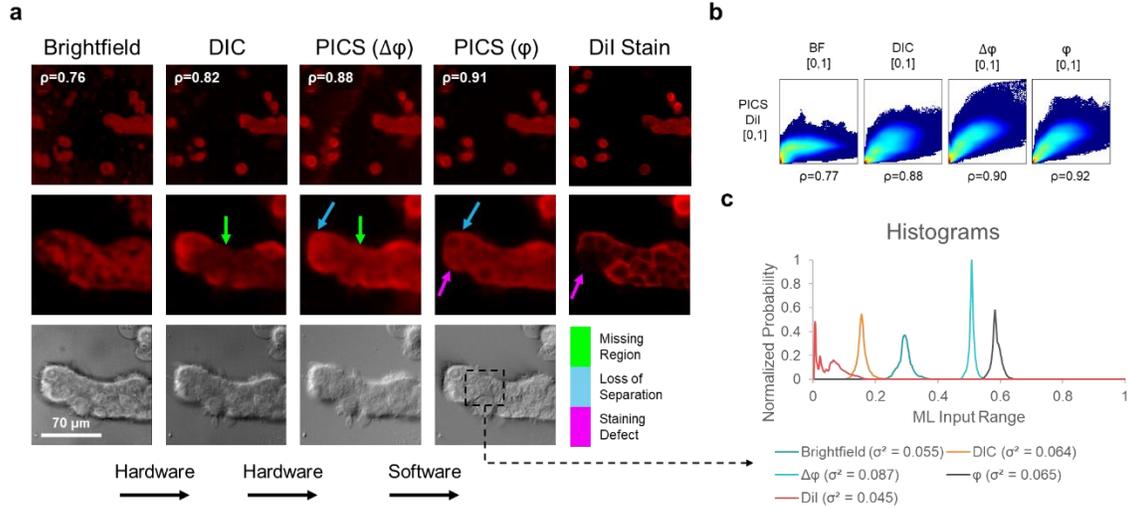

**Supplementary Figure 5: Quantitative phase information improves machine learning performance. a,** To compare the performance of quantitative phase imaging with conventional microscopy, we trained on different stages of the GLIM reconstruction process. Here we take the brightest GLIM frame corresponding to the least interferometric contrast as a brightfield image. DIC denotes the extinction mode frame, Δφ is the GLIM image before integration, and φ is the GLIM image after integration. As a computational experiment we train our U-Net based neural architecture on a subset of DiI images (20x/0.8), for a limited number of epochs, with the same training rate. The performance of the brightfield network is particularly poor, with an improvement when training on the DIC frame. When we introduce phase shifting (Δφ), we isolate the pure phase information resulting in a further improvement in performance (green arrows). This is especially true for at denser portions of the sample where multiple scattering contributes to unwanted amplitude information. While the ground truth image may appear sharper, the PICS neural network was able to pick up cells even when those were not fully represented due to inherent staining defects (pink arrows). **b,** To compare modalities we performed a Pearson correlation across the entire test data set, comparing the measured fluorescence to the computed fluorescent signal, showing that integrated GLIM (φ) has the closest match to the actual fluorescence image. **c,** To investigate the origins of these differences we plot a histogram of the image over a non-empty portion of the sample (dashed black box). When the variance inside this region is used as a contrast metric we note that comparably similar standard deviations (compare DIC at 0.064 to φ at 0.065) lead to different qualitative performances. This result suggests that the difference in performance cannot be erased by simply scaling the data, rather, they are fundamental to the image formation process.

**Supplementary Figure 6**

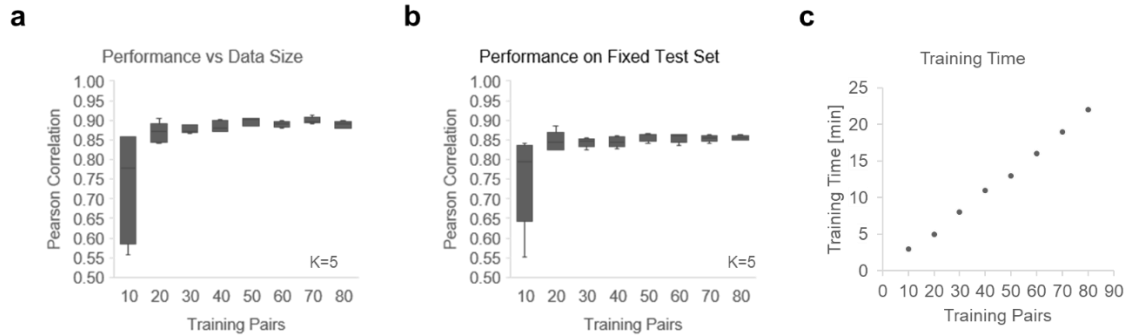

**Supplementary Figure 6: U-Net reaches asymptotic performance with a small number of training pairs.** To better understand the effects of data size on performance we conducted a numerical experiment where we trained PICS networks with an increasing number of QPI-FL training pairs. As per our convention, each "pair" consists of three focus levels, so that with 80 training pairs, we used 240 images for training, 48 for validation, and 174 for the final test group (see Supplementary Table 1). To account for differences in image selection, we perform k-fold validation (k=5), essentially training the network five times for each data set size. **a,** The performance of this network is calculated by looking at the Pearson correlation between the digital and actual fluorescent images. We note that performance becomes asymptotic, hinting that the network is fully trained after approximately 30 pairs. **b,** Each network previously trained is evaluated on an additional 58 fluorescent-phase pairs (174 phase and fluorescence images) that were not used during training. That is to say, we do no vary the test set within each k-fold. We note that this performance also becomes asymptotic after approximately 30 pairs indicating that learning the training corpus has a strong correlation to learning the transformation for unseen data. Looking at the difference in performance within k-fold validation (performance of folds within training pair 10 or 20), we note that some training pairs are substantially more performant, and this performance translates to the unseen test data. **c,** Average time to train a single fold on a single node of the HAL Cluster (NCSA).

**Supplementary Figure 7**

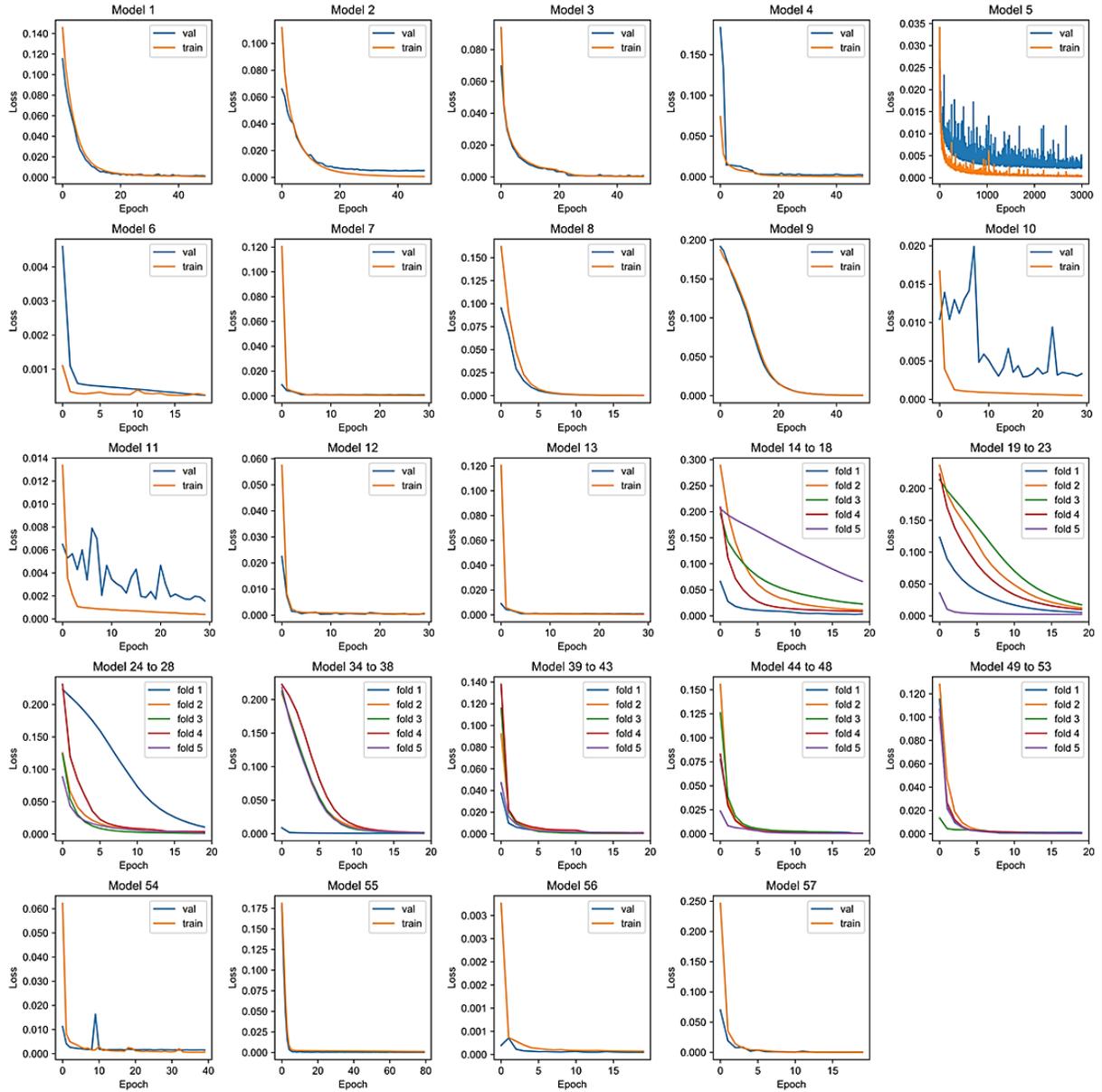

**Supplementary Figure 7: Training Plot for 57 Neural Networks.** To verify that the neural networks converged, we plot the loss on training and validation data after each epoch. A small difference between the two curves indicates that our models do not overfit. A simplified version of this plot is shown for the cross-validated networks used in Supplementary Fig. 6 showing only training loss. We note there is a substantial difference in convergence within the folds used for cross-validation, hinting that some training pairs are easier to learn than others.

## Supplementary Figure 8

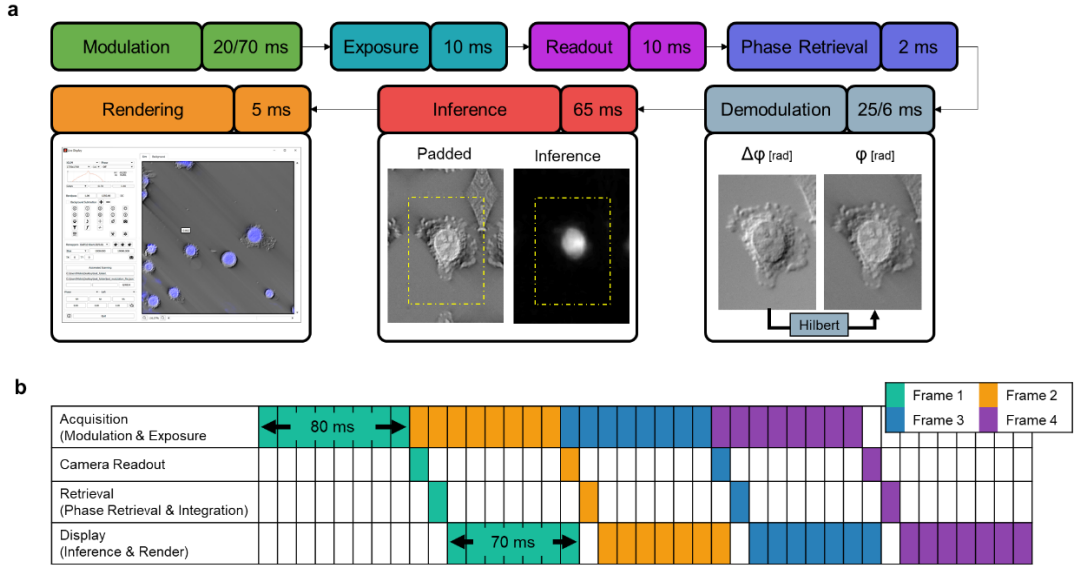

**Supplementary Figure 8: Real-Time acquisition of digital stains from label-free images.**
**a,** In SLIM and GLIM, the acquisition process begins by introducing a controlled modulation which is allowed some time to stabilize (20 ms on SLIM, 70 ms on GLIM). In this work, we acquire full camera frame sizes at minimal exposure (10 ms exposure, 10 ms readout). Phase retrieval is comparably quick (2 ms, 2070 GTX Super, NVIDIA). Phase images typically require further demodulation to correct for system-specific imaging artifacts. In SLIM we perform halo removal to correct for spatial incoherence (25 ms), while GLIM images are integrated along the direction of the DIC shear (6 ms). To avoid edge artifacts, we perform GPU based inference (65 ms) on a larger, mirror padded version of the image, followed by rendering (6 ms). **b,** These steps are performed in parallel to optimally overlap computation with acquisition. As the computation is typically quicker than the acquisition, during real-time operation all reconstruction steps are performed every time a new frame is received. In GLIM the effective frame rate is limited by the modulation of the variable retarder, resulting in one phase image every 80 ms. In practice, it is possible to achieve faster performance, by using high performing graphics card or faster modulators.

**Supplementary Figure 9**

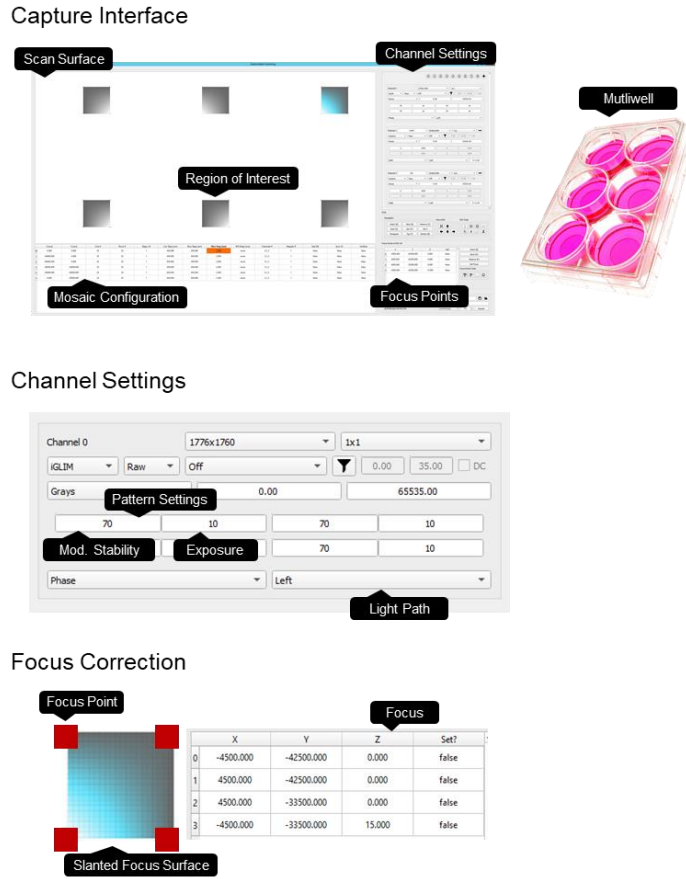

**Supplementary Figure 9: To successfully digitize multiwells we develop a graphic interface that produces a list of acquisition events that are then processed by the acquisition software.** Our capture user interface presents the multiwell as "regions of interest" (rectangles) that have associated "focus points". The interface configures the dimensions of the volume and mosaic parameters such as the number of tiles and steps. The focus points correct for defocus in the sample (mostly due to mounting), and the scan is performed offset to the estimated tilt. In addition to configuring fluorescence acquisition, our interface contains phase imaging specific parameters such as the modulator stabilization times and exposure for each pattern used to reconstruct the phase image.

**Supplementary Figure 10**

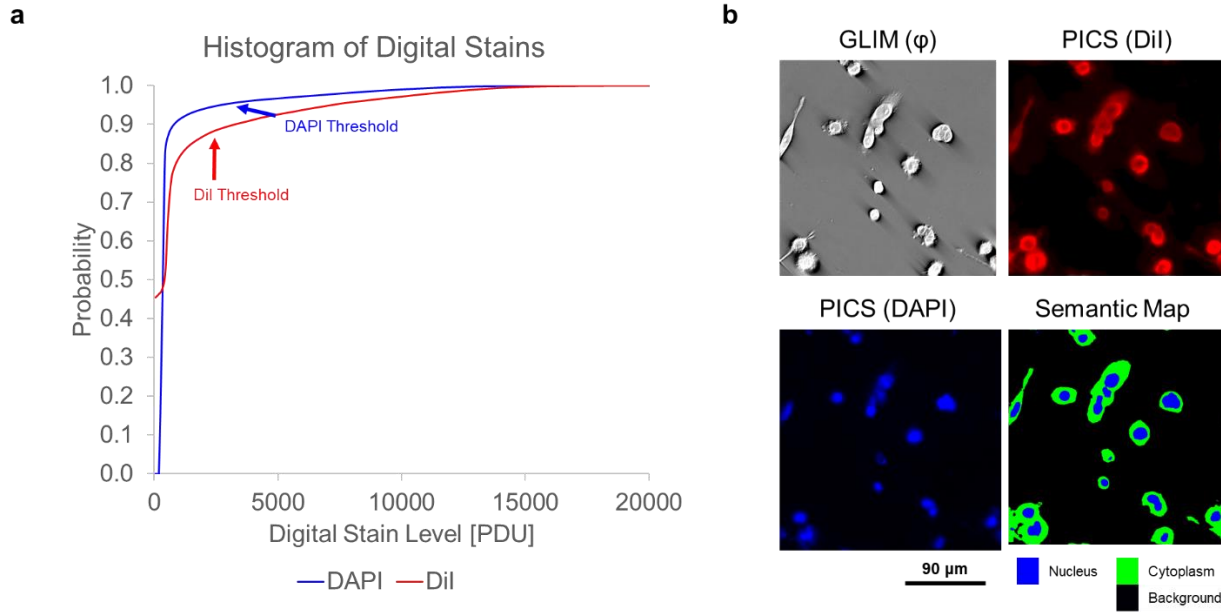

**Supplemental Figure 10: Semantic segmentation map from digital stains. a,** Digitally stained images were binarized to discriminate between stain and background by analyzing the cumulative sum of a histogram for a representative image. It was found that the change in inflection of the cumulative histogram of fluorescence intensity values served as a good threshold marker. Intuitively this change in inflection indicates when the histogram switches from tracking the background to the sample. In this work, we used the same binarization thresholds for all training pairs. **b,** Thresholded images were combined into a semantic segmentation map, by labeling all pixels with the PICS-DAPI binary mask blue, all pixels that had a PICS-DiI mask but were not blue as green (cytoplasm), and all pixels not labeled as either of the two as black (background).

**Supplementary Figure 11**

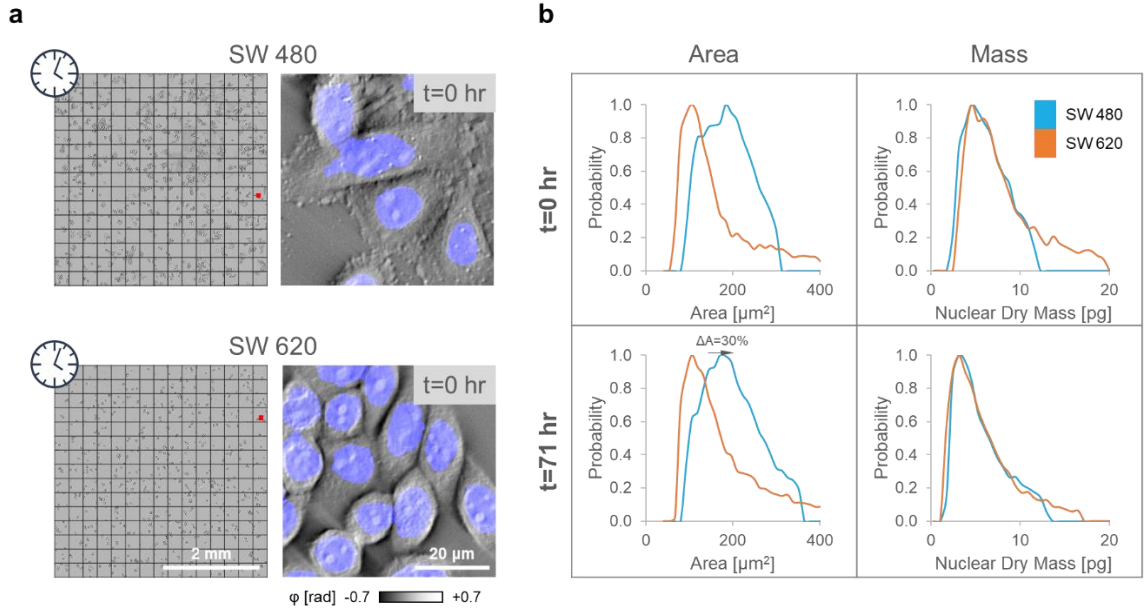

**Supplemental Figure 11: Time-lapse monitoring of nuclear dry mass and area for SW 480 and SW 620 subclones. a,** SW cells were images for 72 hours with a multiwell scan acquired every two hours. The semantic segmentation map from PICS was used to generate markers and ridgelines to perform instance segmentation using a watershed-based approach. **b,** instance segmentation on the cellular nucleolus shows that while SW 620 (metastatic) has somewhat smaller nuclei (ΔA=30%) total nuclear dry mass remains relatively consistent between SW 480 and SW 620.

**Supplementary Figure 12**

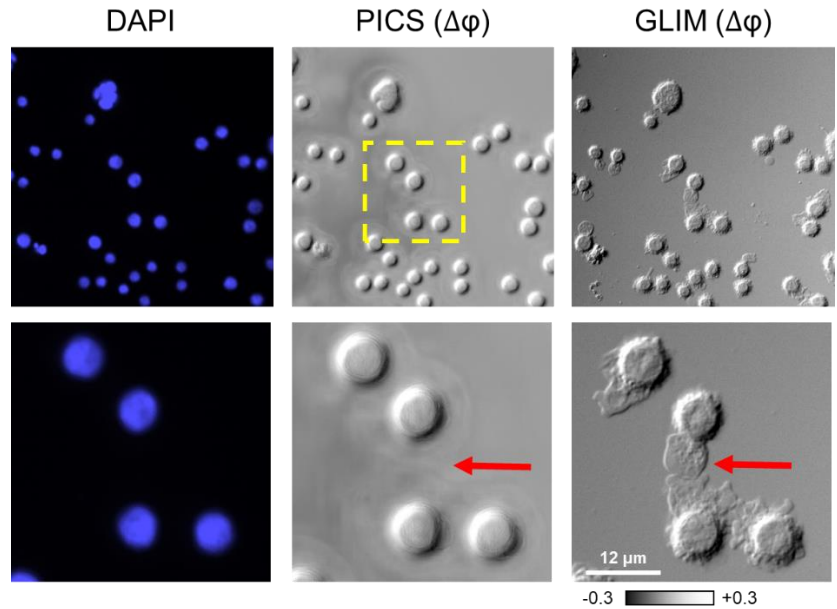

**Supplemental Figure 12: Fluorescence microscopy provides limited information compared to scattered light imaging.** While scattered light GLIM images can be used to produce fluorescence equivalents, attempting to do the reverse, going from DAPI to phase images has substantially worse performance missing structural details that would lead to a substantial underestimation of the cell's area (red arrows).

**Supplementary Table 1**

| ID*** | Appearance | Modality | Objective | Stain | Cell Line** | Training Pairs | Validation Pairs | Test Pairs | Learning Rate | Epochs | Pearson Correlation |
|---|---|---|---|---|---|---|---|---|---|---|---|
| 1 | Fig 3 | GLIM | 5x/0.08 | DAPI | SW | 48 | 6 | 6 | 5e-5 | 50 | 0.75 |
| 2 | Fig 3 | GLIM | 10x/0.30 | DAPI | SW | 48 | 6 | 6 | 5e-5 | 50 | 0.87 |
| 3 | Fig 3 | GLIM | 20x/0.80 | DAPI | SW | 48 | 6 | 6 | 5e-5 | 50 | 0.89 |
| 4 | Fig 3 | GLIM | 63x/1.40 | DAPI | SW | 48 | 6 | 6 | 5e-5 | 50 | 0.94 |
| 5 | Fig 6 | GLIM | 63x/1.40 | DAPI | HepG2 | 16324 | 64 | 32 | 1e-4 | 3000 | 0.97 |
| 6 | Fig S4 | GLIM | 20x/0.80 | DAPI | SW | 830 | 42 | 10 | 1e-4 | 20 | 0.93 |
| 7 | Fig S4 | GLIM | 20x/0.80 | DiI | SW | 705 | 87 | 90 | 5e-5 | 30 | 0.94 |
| 8 | Fig S4 | SLIM | 10x/0.30 | DAPI | SW | 210 | 30 | 30 | 1e-4 | 20 | 0.92 |
| 9 | Fig S4 | SLIM | 20x/0.30 | DAPI | CHO | 48 | 6 | 6 | 5e-5 | 50 | 0.86 |
| 10 | Fig S5 | BF | 20x/0.80 | DiI | SW | 705 | 87 | 90 | 5e-5 | 30 | 0.51 |
| 11 | Fig S5 | DIC | 20x/0.80 | DiI | SW | 705 | 87 | 90 | 5e-5 | 30 | 0.87 |
| 12 | Fig S5 | GLIM ($\Delta\varphi$) | 20x/0.80 | DiI | SW | 705 | 87 | 90 | 5e-5 | 30 | 0.92 |
| 13 | Fig S5 | GLIM | 20x/0.80 | DiI | SW | 705 | 87 | 90 | 5e-5 | 30 | 0.94 |
| 14-18 | Fig S6 | GLIM | 20x/0.80 | DAPI | SW | 30 | 6 | 174 | 1e-4 | 20 | *0.51\** |
| 19-23 | Fig S6 | GLIM | 20x/0.80 | DAPI | SW | 60 | 12 | 174 | 1e-4 | 20 | *0.65\** |
| 24-28 | Fig S6 | GLIM | 20x/0.80 | DAPI | SW | 90 | 18 | 174 | 1e-4 | 20 | *0.82\** |
| 29-33 | Fig S6 | GLIM | 20x/0.80 | DAPI | SW | 120 | 24 | 174 | 1e-4 | 20 | *0.73\** |
| 34-38 | Fig S6 | GLIM | 20x/0.80 | DAPI | SW | 150 | 30 | 174 | 1e-4 | 20 | *0.89\** |
| 39-43 | Fig S6 | GLIM | 20x/0.80 | DAPI | SW | 180 | 36 | 174 | 1e-4 | 20 | *0.86\** |
| 44-48 | Fig S6 | GLIM | 20x/0.80 | DAPI | SW | 210 | 42 | 174 | 1e-4 | 20 | *0.90\** |
| 49-53 | Fig S6 | GLIM | 20x/0.80 | DAPI | SW | 240 | 48 | 174 | 1e-4 | 20 | *0.88\** |
| 54 | Fig S11 | GLIM | 20x/0.80 | DAPI | SW | 390 | 45 | 45 | 1e-4 | 40 | 0.91 |
| 55 | Fig S11 | GLIM | 20x/0.80 | DiI | SW | 3000 | 300 | 300 | 1e-5 | 80 | 0.83**** |
| 56 | Fig S12 | GLIM | 20x/0.80 | DAPI | SW | 705 | 87 | 90 | 5e-5 | 20 | 0.78 |
| 57 | Vid S3 | SLIM | 10x/0.30 | DAPI | SW | 210 | 30 | 30 | 1e-4 | 20 | 0.80 |

\* Averaged performance across all k models for that k-fold cross-validation experiment

\*\* SW network was trained on both SW480 and SW620

\*\*\* Corresponding to IDs in supplementary Fig. 7

\*\*\*\* Not used for analysis

**Video 1**

Co-localized acquisition of GLIM and DAPI data for PICS training (20x/0.8, SW cells).

**Video 2**

Real-time GLIM and PICS-DAPI (20x/0.8, SW cells).

**Video 3**

Real-time SLIM and PICS-DAPI (10x/0.3, CHO cells).

**Video 4**

Time-lapse GLIM and PICS imaged over seven days (20x/0.3, SW Cells).